\documentclass[12pt,preprint]{aastex}

\usepackage{epsfig}
\usepackage{graphicx}
\usepackage{epstopdf}
\usepackage{amsmath}
\usepackage{amssymb}
\usepackage{appendix}

\def\be{\begin{equation}}
\def\ee{\end{equation}}
\def\ba{\begin{eqnarray}}
\def\ea{\end{eqnarray}}
\def\go{\mathrel{\raise.3ex\hbox{$>$}\mkern-14mu
             \lower0.6ex\hbox{$\sim$}}}
\def\lo{\mathrel{\raise.3ex\hbox{$<$}\mkern-14mu
             \lower0.6ex\hbox{$\sim$}}}

\begin{document}

\title{Magnetar Giant Flares in Multipolar Magnetic Fields --- \\
II. Flux Rope Eruptions With Current Sheets
}

\author{Lei Huang\altaffilmark{1,3,4} and Cong Yu\altaffilmark{2,4}}
\altaffiltext{1}{Key
Laboratory for Research in Galaxies and Cosmology, Shanghai
Astronomical Observatory, Chinese Academy of Sciences, Shanghai,
200030, China; {\tt muduri@shao.ac.cn}}
\altaffiltext{2}{Yunnan Observatories, Chinese Academy of Sciences, Kunming,
650011, China; {\tt cyu@ynao.ac.cn}}
\altaffiltext{3}{Key Laboratory of Radio Astronomy, Chinese
Academy of Sciences, China. }
\altaffiltext{4}{Key
Laboratory for the Structure and Evolution of Celestial Object,
Chinese Academy of Sciences, Kunming, 650011, China; }

\begin{abstract}
We propose a physical mechanism to explain giant flares and radio
afterglows in terms of a magnetospheric model containing both a
helically twisted flux rope and a current sheet (CS). With the
appearance of CS, we solve a mixed boundary value problem to get
the magnetospheric field based on a domain decomposition method.
We investigate properties of the equilibrium curve of the flux rope 
when the CS is present in background multipolar fields. In
response to the variations at the magnetar surface,
it quasi-statically evolves in stable equilibrium states. 
The loss of equilibrium occurs at a critical
point and, beyond that point, it erupts catastrophically. New
features show up when the CS is considered. Especially, we find
two kinds of physical behaviors, i.e., catastrophic state
transition and catastrophic escape. Magnetic energy would be
released during state transitions. The released magnetic energy is
sufficient to drive giant flares. The flux rope would go away from
the magnetar quasi-statically, which is inconsistent with the
radio afterglow. Fortunately, in the latter case, i.e., the
catastrophic escape, the flux rope could escape the magnetar and
go to infinity in a dynamical way. This is more consistent with
radio afterglow observations of giant flares. We find that the
minor radius of flux rope has important implications for its
eruption. Flux ropes with larger minor radius are more prone to
erupt. We stress that the CS provides an ideal place for magnetic
reconnection, which would further enhance the energy release
during eruptions.
\end{abstract}


\section{Introduction}

Soft gamma-ray repeaters (SGRs) and anomalous X-ray pulsars (AXPs) 
are widely recognized to be magnetars $-$ highly magnetized
isolated neutron stars \citep{Maze79,DT92,Kouv98}.
Compared to radio pulsars, the magnetic fields of magnetars are at least $10^2-10^3$ times stronger, 
i.e., up to $10^{14}-10^{15}$G.
Persistent and bursting emissions from
these sources are well explained by the dissipation of their ultra-strong magnetic fields
\citep{MS95,HK98,TLK02,GKW02}.
In addition to persistent emissions and short duration bursts, the
most spectacular and perplexing phenomenon associated with
magnetars is the giant flare.
The peak luminosity during this event is over $10^6$ times higher than
the Eddington luminosity of a typical neutron star,
releasing total energy of $\sim 10^{44}-10^{46}$erg \citep{WT06,Mere08}.
How the magnetic energy of giant
flare is stored and released still remains a puzzling question and
is now under extensive debate \citep{TD01,Lyut06,GH10,Yu12,Parf13,Link14,Meng14,Taka14}.

Two discrepant physical models are put forward to explain the
giant flare. The main distinction lies in the location of the magnetic energy build-up
prior to the eruptive outbursts:
in the crust \citep{TD01} or in the magnetosphere \citep{Lyut06}.
In the crust model,  the giant flare is
considered to be initiated by a sudden untwisting of the neutron
star's internal magnetic field. The resulting magnetic stress
variations lead to fracture of the crust and
trigger violent outbursts \citep{TD01}. However, recent
calculation shows that crust model has certain difficulties in
explaining giant flares \citep{LL12,Link14}.
Alternative models, i.e., magnetospheric models have been
established to explain the magnetic energy storage and eruptive
outbursts related to magnetar giant flares
\citep{Yu11b,Yu12,HY14}.
In such models, the magnetic energy is thought to accumulate
quasi-statically in the magnetosphere with the neutron star
surface gradual variations.
Once the system evolves to a critical state, the flux rope would
erupt suddenly on a rapid dynamical timescale, which is consistent
with the short rise timescale of giant flare, $\sim0.25$ms
\citep{Palm05}. Note that the energy release timescale of the
giant flare is much shorter than the energy accumulation
timescale. Such a distinctive timescale contrast brings about an
interesting question. How could the long timescale energy buildup
process lead to an abrupt 
energy release on a dynamical timescale? 
To answer this question, we have attempted to construct
a catastrophic flux rope eruption model to explain the initiation
of the giant flare \citep{Yu12,HY14}. The authors showed that, with
the quasi-static variations at the magnetar surface, either crust
motions or flux injections, the flux rope would
experience a quasi-static evolution. 
Once approaching a critical height, the flux rope would erupt due to the loss of equilibrium.
The catastrophic feature gives a natural solution to the above question. 
The magnetic energy is built up gradually in magnetosphere before reaching the critical point.
Upon the critical point where the flux rope is ready to erupt,
the magnetic energy accumulated in previous evolution is expected to be released on a dynamical timescale.
In these flux rope eruption models,
what we do know is that there indeed exists a critical point along
the equilibrium curve and the flux rope would lose its equilibrium
and erupts when it reaches this critical point. However, we are
still not sure about the flux rope's subsequent behavior after the
catastrophic eruption, which could be either quasi-static or
dynamic responses.

It is well known that magnetic reconnection plays a crucial role
in the magnetic energy dissipations \citep{PF00,GH10,MU12,LT12},
which is important for both magnetars and rotation-driven pulsars, especially for the Crab Nebula flares
in recent observation \citep[e.g.,][]{SA12}. 
The secondary plasmoid instability is also expected to take place in the CS \citep{HB12}.
The plasmoid ejection induced reconnection
process could probably lead to fast variabilities.
It has been hypothesized that a current sheet, where magnetic
reconnection could naturally take place, can be generated in the
magnetosphere \citep{Lyut06,GH10}.
In our flux rope eruption model, detailed calculations show that {\bf when} 
the flux rope becomes unstable, the CS is indeed created in the magnetosphere \citep{YH13}.
The formation of current sheet and subsequent magnetic
reconnection would lead to rapid magnetic field dissipation 
and powerful release of magnetic energy. The formation of current
sheet results in a considerable topological change in the magnetic
field configuration. When a current sheet is present, the global structure of 
magnetosphere becomes a rather complex mixed boundary
value problem. We have figured out a numerical scheme to construct
the magnetospheric field with a current sheet. Details will be
further discussed in the following sections.

The pulse profile of the 1998 Aug 27 event observed from SGR 1900+14 indicates that
in the vicinity of star, 
the magnetic geometry was far more complicated compared to a dipole 
and higher multi-poles may be involved \citep{Fero01}. During the
birth of magnetars, electric currents are formed within the
magnetar interior. These currents slowly push out
and generate magnetic active regions, shown as multipolar regions, on the magnetar surface.
These multipolar magnetic fields may also well affect the radiative
emission properties of the magnetars \citep{Pava09}. In
\citet{HY14}, hereafter Paper I, we study the flux rope eruption
model in the multipolar background field.
The energetics of this model is carefully investigated. It is
found that the accumulated free magnetic energy in this model
could be sufficient to drive a giant flare, which releases energy
about $1\%$ of the total magnetic energy of neutron star in
observations \citep{WT06,Mere08}.
However, we did not consider effects of the current sheet in
Paper I. In this paper, we extends our previous calculation by
including the effects of current sheet. New features introduced by
the current sheet are further investigated and their astrophysical
implications are discussed in this paper.

The magnetic energy release associated with catastrophic state
transitions may account for the giant flare itself. But
observations showed that, large scale radio afterglow ($\sim
0.1$pc) were also detected following the giant flares on 1998 Aug 27 from SGR 1900+14
and on 2004 Dec 27 from SGR 1806-20.
The radio afterglow indicates that there are relativistic
plasma escaping from the central magnetar after the violent
outbursts \citep{Gaen05}. The radio afterglow remains a puzzle for
the flux rope eruption model \citep{YH13}. This is because,
after the catastrophic loss of equilibrium, the flux rope would
reach another stable equilibrium state and evolve quasi-statically
later on.
However, the observed radio afterglow implies dynamically escaping
plasma, which is hard to be explained by our previous model.
Fortunately, we find an interesting behavior of the flux rope -
catastrophic escape. If the flux distribution on the surface involves higher order multipoles, 
upon the loss of equilibrium point,
the flux rope cannot obtain a new equilibrium state, 
but escape dynamically to infinity.
Catastrophic escape of the flux rope may naturally explain both
the giant flare and the radio afterglow of SGR1806-20 \citep{Gaen05}.

This paper is organized 
as follows. In Section 2, we introduce the model of multipolar magnetosphere
containing a flux rope as well as a current sheet accompanied.
The numerical method to handle both the flux rope and the current
sheet in a global solution of magnetosphere is described in
details in Section 3. Examples of the equilibrium curve of the
system and two types of catastrophic behavior after loss of equilibrium 
are presented in Section 4. In Section 5, the energetics of our model and
the effects of the flux rope's minor radius
on flux rope eruptions are discussed.
Conclusions and discussions are provided in Section 6.

\section{Force-Free Magnetosphere with Both Flux Rope and Current Sheet}
The magnetosphere is magnetically dominated and the
pressure and inertia of the plasma can be neglected \citep{TLK02}.
This provides a rather neat and clean description of
the magnetosphere, in which the highly magnetized plasma can be 
assumed in a force-free state, i.e., $\mathbf{J} \times
\mathbf{B} = 0$. The model here distinguishes itself from that in
Paper I in that an additional component, the current sheet, is included.

\subsection{Stationary Axisymmetric Force-Free Magnetosphere}
As in paper I, we consider a toroidal and helically twisted flux
rope embedded in magnetosphere. 
The magnetic fields inside flux rope and those outside
it should be treated separately. Inside flux rope, the
magnetic fields are considered to be generated by a net current
$I$ carried by the flux rope.
We adopt the force-free solution \citep{Lund50} to represent the
magnetic field and current density distribution inside flux rope.
The height of the toroidal flux rope is denoted as a major radius $h$,
and the size of it is denoted as a minor radius $r_0$ \citep{Yu12,HY14}.
Note that the magnetic twist is well-confined inside the flux rope.
Our model is different from the global magnetic twist model considered
in \citet{TLK02} and \citet{Belo09}.
The axisymmetric magnetic field $\mathbf{B}$ outside the flux rope
can be expressed in spherical coordinates $(r,\theta,\phi)$ as
follows,
\begin{equation}\label{BrBtheta}
\mathbf{B} = - \frac{1}{r^2} \frac{\partial\Psi}{\partial\mu}\
\hat{\mathbf{e}}_r \ -\frac{1}{r\sin\theta}
\frac{\partial\Psi}{\partial r}\ \hat{\mathbf{e}}_\theta  \ ,
\end{equation}
where $\mu = \cos\theta$ and $\Psi = \Psi(r,\mu)$ is the magnetic
stream function. Note that $\hat{\mathbf{e}}_r$ and
$\hat{\mathbf{e}}_\theta$ are unit vectors respectively in radial and
latitudinal direction.
We express the force-free condition as the inhomogeneous Grad-Shafranov (GS) equation. 
Explicitly,
\begin{equation}\label{GS}
\frac{\partial^2\Psi}{\partial r^2} + \frac{(1-\mu^2)}{r^2}\
\frac{\partial^2\Psi}{\partial\mu^2}\ =\ - r\sin\theta\
\frac{4\pi}{c}J_\phi \ ,
\end{equation}
where $c$ is the speed of light. On the right hand side, $J_\phi$
represents the current density induced by the toroidal flux rope,
which is treated as a circular ring current of, $J_{\phi}
= (I/h)\delta(\mu)\delta(r-h)$, where, $I$, as mentioned above,
designates the electric current carried by the flux rope
\citep{PF00,HY14}.

According to Paper I, when the helically twisted flux rope evolves
to a critical state, it would lead to a catastrophic eruption. A
current sheet would naturally form during the eruption
\citep{YH13}. This current sheet is thought to connect the
magnetar surface and the flux rope. Obviously, the appearance of the current sheet
would change the topology of the magnetospheric field.  We will
focus on new solutions with current sheet to the GS equation in
this paper. The numerical method to handle the current sheet will
be further discussed in the following sections.

\subsection{Mixed Boundary Value Problems for Force-Free Magnetospheres}
Three physical boundary conditions needs to be satisfied to get
the self-consistent global magnetospheric field with a current
sheet. The first boundary condition,
i.e., $|\nabla\Psi| \to 0$ for $r\to\infty$, can be satisfied trivially \citep[see][]{Yu12}.
The second boundary is located at the magnetar surface $r=r_s$.
Following Paper I, multipolar boundary conditions are adopted
at magnetar surface, i.e.,
\begin{equation}\label{BCsurf}
\Psi(r_s,\mu) = \Psi_0 \sigma \Theta(\mu) \ ,
\end{equation}
where $\Psi_0$ is a constant parameter which has dimension of magnetic flux 
and the dimensionless variable $\sigma$ represents the
magnitude of surface magnetic flux. 
Here $\Theta(\mu)$ depicts the angular dependence of the stream
function at the magnetar surface, which must be chosen properly
according to the observations of magnetars. Although it is widely
believed that magnetic field configuration of magnetar is basically a dipole,
observations showed that the geometry of magnetic field involves multipoles
in the vicinity of surface \citep{Fero01}.
To model the multipolar field, 
we assume that the angular dependence of the stream function is contributed by
a dipolar component plus a high order multipolar component. Specifically,
\begin{equation}\label{pb}
\Theta(\mu) = (1 -\mu^2)\ +\ a_1\ (5\mu^2 - 1) (1 - \mu^2) \ ,
\end{equation}
where first term $(1-\mu^2)$ denotes the dipolar component and
the latter term with strength coefficient $a_1$ represents the contribution of high-order multipoles.
Properties of this boundary condition has been discussed in Paper I.
We do not repeat them here.

The third boundary is introduced by the current sheet. The current
sheet is represented by a horizontal thick solid line in Fig. 1.
The current sheet lies at $\theta = \pi/2$, the equatorial plane,
and covers the radial range between $r_s$ and $r_1$, where $r_s$
and $r_1$ stand for the magnetar radius and the tip of the current
sheet, respectively. The existence of the equatorial current sheet
requires a third boundary condition to be fulfilled, which
requires
\begin{equation}\label{boundaryCS}
\Psi(r,0) \equiv \Psi_{\rm cs}   \  ,  \ r_s \leqslant r \leqslant
r_1 \ ,
\end{equation}
where $\Psi_{\rm cs}$ denotes the constant value of stream
function $\Psi$ along the current sheet. The value of $\Psi_{\rm
cs} = \Psi_0 \sigma \Theta_0$, where $\Theta_0 = \Theta(0)$ is the
value of $\Theta(\mu)$ at $\mu=0$. It is clear that $\Theta_0 = 1
- a_1$ according to the surface boundary condition we adopt above.
Due to this additional boundary condition at the current sheet,
the GS Equation (\ref{GS}) must be solved in a non-trivial way numerically.
We have figured out an appropriate numerical scheme to get the global
configuration of the magnetospheric field, which will be further
discussed in the following section.

\section{Domain Decomposition --- Numerical Treatment about Current Sheets }

The most distinctive feature of our model is that, in addition to
the flux rope with internal helically twisted magnetic field inside,
a current sheet is incorporated.
In this case, the solution to the GS equation constitutes a problem with mixed boundary value.
The current sheet provides another boundary to be considered apart from the magnetar surface.
Our numerical strategy is the domain decomposition.

\begin{figure}
\includegraphics[scale=0.9]{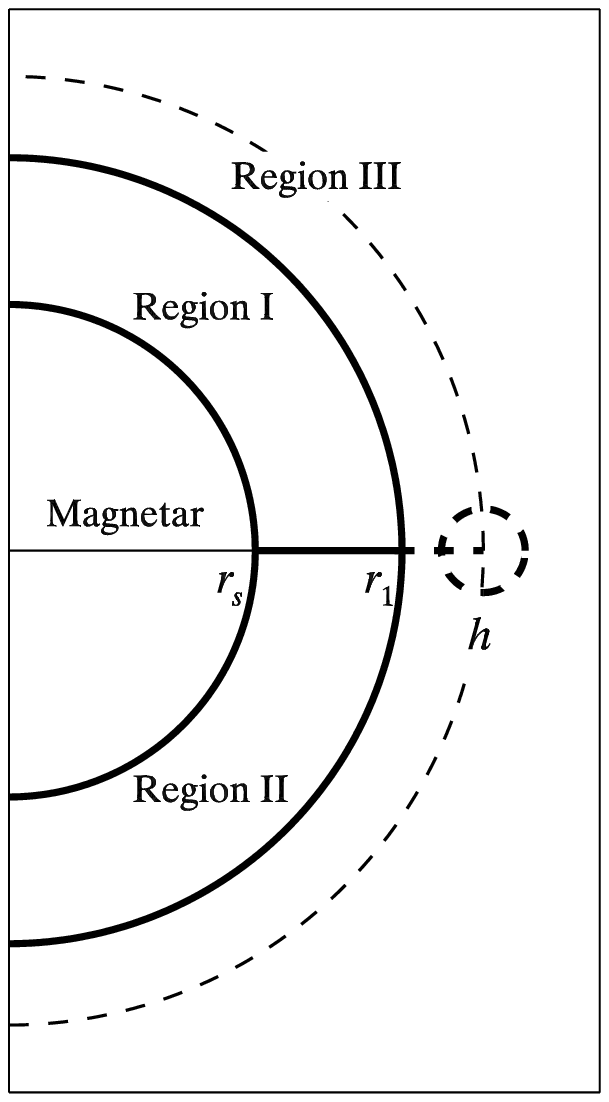}
\caption{\label{diag} Schematic illustration of magnetosphere
containing both a flux rope and a current sheet. The radius of
magnetar is denoted by $r_s$. The current sheet is designated by
the thick horizontal line at the equatorial plane, the upper tip
of which is denoted by $r_1$. The current sheet separates region I
and II. The flux rope, located in region III, is designated as
thick dashed circle at the equator. The height of the flux rope is
denoted by $h$.
 }
\end{figure}

In Fig.\ref{diag}, we give a schematic illustration of the magnetosphere.
The inner thick solid semi-circle in radius of $r_s$ represents the magnetar surface
and the thick dashed circle denotes the flux rope.
The thin dashed semi-circle with radius,
$h$, which goes through the center of flux rope, represents its
height measured from the magnetar center. The current sheet, lying
on the equatorial plane, is shown as thick horizontal solid line.
Note that the lower tip of the current sheet is connected to the
magnetar surface. The upper tip is located at $r_1$.
It is obvious that the current sheet divides the global force-free
magnetosphere into three regions, labelled as Region I, II, and
III, respectively. The flux rope lies in Region III and the
current sheet separates Region I and Region II. To be more
specific, Region I covers the region $r_s \leqslant r\leqslant
r_1$ and $0 \leqslant \theta \leqslant \pi/2$.
Region II is symmetric to Region I and covers the region
$r_s\leqslant r\leqslant r_1$ and $\pi/2 \leqslant \theta
\leqslant \pi$. Region III indicates the area $r\geqslant r_1$ and
$0\leqslant\theta\leqslant\pi$. We can readily construct separate
local solutions in these different regions. However, we have to
meld all the local solutions in different regions together to
establish a self-consistent global magnetosphere.

\subsection{Local Solutions in Different Regions}
The general solution for the stream function in Region I can be
written as \citep{Yu12}
\begin{equation}\label{psiI}
\Psi_\mathrm{I}(r,\mu) = \Psi_{\rm cs} (1-\mu) + \sum_{k=1}^\infty
\left( a_{2k}r^{2k+1} + b_{2k}r^{-2k} \right)\ \left[
\frac{P_{2k-1}(\mu)-P_{2k+1}(\mu)}{4k+1} \right] \ ,  \ 0
\leqslant\mu \leqslant 1 \ ,
\end{equation}
where $\Psi_{\rm cs}$ is already defined in Equation (\ref{boundaryCS}),
$P_{2k-1}(\mu)$ and $P_{2k+1}(\mu)$ are Legendre polynomials,
and the constant coefficients $a_{2k}$ and $b_{2k}$ will be specified 
following the numerical strategy in Appendix A.
It is worthwhile to note that in the
above equation, only Legendre polynomials with odd orders
are involved.
Based on the above explicit solution, it is readily to know that
current sheet boundary condition, Equation (\ref{boundaryCS}), can
be trivially satisfied. When it comes to the magnetar surface
boundary condition Equation (\ref{BCsurf}), since the
magnetosphere is symmetric 
to the equatorial plane,
we only need to consider half of the magnetar's surface, i.e., the
northern hemisphere. In other words, we only consider Equation
(\ref{BCsurf}) for $0 \leqslant\theta\leqslant\pi/2$. As a result,
this boundary condition can be cast in the following explicit form,
\begin{equation}
\Psi_{\rm I}(r_s, \mu) =  \Psi_0 \sigma \Theta(\mu) \ , \ 0
\leqslant\mu \leqslant 1 \ ,
\end{equation}
According to the orthogonality of Legendre polynomials, the above
equation means that  coefficients $a_{2k}$'s and $b_{2k}$'s
satisfy the following relation\footnote{For numerical
conveniences, we scale magnetic flux by $\Psi_0$, current by
$I_0\equiv\Psi_0 c/r_s$ and all lengths by the neutron star radius
$r_s$ throughout this paper. Obviously, $r=1$ at the neutron star
surface.},
\begin{equation}\label{a2kb2k}
a_{2k} + b_{2k} = T_{2k} \ ,
\end{equation}
where $T_{2k}$ is explicitly specified in accordance with the
magnetar surface flux distribution $\Theta(\mu)$ in Equation
(\ref{pb}) as
\begin{equation}
T_{2k} = (4k+1)\ \Psi_0 \sigma \int_0^1 \left[  \Theta(\mu) -
\Theta_0 (1 - \mu) \right]\ \frac{dP_{2k}(\mu)}{d\mu}\ d\mu\ .
\end{equation}
Since $a_{2k}$ and $b_{2k}$ are closely related by Equation
(\ref{a2kb2k}), once the coefficients $a_{2k}$'s are determined,
$b_{2k}$ can be easily identified.
Because the solution in Region II is symmetric to that in Region
I, we do not reiterate on the explicit form of $\Psi_{\rm
II}(\mu)$. In the computational domain of Region III where the
flux rope is embedded, the general solution of the magnetic stream
function is of the following form (Yu 2012)
\begin{equation}\label{psiIII}
\Psi_\mathrm{III}(r,\mu) = \sum_{i=0}^\infty \left[
c_{2i+1}R_{2i+1}(r)+d_{2i+1}r^{-2i-1} \right]\
\left[\frac{P_{2i}(\mu)-P_{2i+2}(\mu)}{4i+3}\right]\ ,
\end{equation}
where the piecewise continuous function $R_{2i+1}(r)$ is defined
as in Paper I, the coefficients $c_{2i+1}$'s can be determined in
terms of the current carried by the flux rope, and $d_{2i+1}$'s
are coefficients to be fixed in terms of the matching condition
discussed in the following section. It is interesting to note that
only Legendre polynomials with even orders, $P_{2i}(\mu)$ and
$P_{2i+2}(\mu)$, are involved in the above equation. This is
different from the fact that only Legendre polynomials with odd orders
appear in Equation (\ref{psiI}).

A global solution of magnetosphere can be constructed
once local solutions, $\Psi_{\rm I}$, $\Psi_{\rm II}$, and $\Psi_{\rm III}$,
in three different computational Regions, I, II, III, are
obtained. In other words, once the coefficients $a_{2k}$'s,
$b_{2k}$'s, $c_{2i+1}$'s, and $d_{2i+1}$'s are self-consistently
determined, a global magnetosphere with both flux rope and current
sheet can be naturally established. In the following section, we
will discuss the numerical scheme to self-consistently determine
these coefficients in greater details.

\subsection{Matching Conditions for Different Regions}
Up to now, we have explicitly obtained the local solutions,
$\Psi_{\rm I}$, $\Psi_{\rm II}$, and $\Psi_{\rm III}$, in Region
I, II, and III, respectively. These local solutions have
distinctive forms in these regions. To meld them into a consistent
global solution, we need to seamlessly match these local
solutions. Physically speaking, the global magnetic field should
be continuous throughout the entire magnetosphere, which is
trivially satisfied within each regions. However, this is not the
case across the interface between different regions.

The global solution must satisfy the following constraints to smoothly connect the magnetic field 
between Region I (II) and Region III, viz.,
\begin{eqnarray}
    &\ & \left\{\begin{array}{cc}
        {\bf B}_{\mathrm{I}}(r_1,\mu)={\bf B}_{\mathrm{III}}(r_1,\mu)\ ,&0\le\mu\le1 \\
        {\bf B}_{\mathrm{II}}(r_1,\mu)={\bf B}_{\mathrm{III}}(r_1,\mu)\ ,&-1\le\mu\le0\\
    \end{array}  \right.\ .
\end{eqnarray}
Because of the symmetry of the magnetosphere with respect
to the equatorial plane, it is sufficient to consider the relevant
equation between Region I and Region III for practical
calculations.
Once the solution in the northern-hemisphere (Region I,
$0\le\mu\le1$) is obtained, the solution in the
southern-hemisphere (Region II, $-1\le\mu\le0$) can be easily
specified according to the symmetry properties of the
magnetosphere. Note that the above equation in the northern
hemisphere ($0\le\mu\le 1$) is a vector equation, which can be
written, according to Equation (\ref{BrBtheta}), as
\begin{equation}\label{matchBr}
\frac{\partial\Psi_{\rm I}}{\partial\mu}\bigg|_{(r_1,\mu)} =
\frac{\partial\Psi_{\rm III}}{\partial\mu}\bigg|_{(r_1,\mu)} \ ,
\end{equation}
\begin{equation}\label{matchBth}
\frac{\partial\Psi_{\rm I}}{\partial r}\bigg|_{(r_1,\mu)} =
\frac{\partial\Psi_{\rm III}}{\partial r}\bigg|_{(r_1,\mu)}  \ ,
\end{equation}
where $\mu$ is in the range of $\mu\in[0,1]$. It is found that the
numerical implementation of the above matching condition is rather
complicated \citep{YH13}. To avoid digression of the main theme of
this paper, we will discussed detailed numerical matching
procedures in Appendix A.

\subsection{Y-point Condition at Current Sheet Tip}

The current sheet tip ($r=r_1,\mu=0$), is located on the
equatorial plane and separates Region III and Region I (or II).
Note that according to Equation (\ref{BrBtheta}) and Equation
(\ref{psiI}), the value of $B_\theta(r\to r_1^-,\mu=0)$ is always
zero in Region I, since all odd-term Legendre polynomials
$P_{2k+1}(\mu=0)=0$. To keep the magnetic field continuous at the
current sheet tip, $B_{\theta}(r\to r_1^{+}, \mu=0)$ in Region III
must be also zero. This condition is called Y-point condition,
which explicitly reads,
\begin{equation}\label{Ypoint}
B_\theta(r\to r_1^+, \mu = 0) = 0 \ .
\end{equation}
With the magnetic flux on magnetar surface $\sigma$ fixed, to
construct a global magnetosphere model we have to specify another
three parameters, i.e., the dimensionless current $J$, the flux
rope major radius $h$, and the length of the current sheet tip
$r_1$. However, this Y-point condition indicates that the location
of current sheet tip, $r_1$, can not be chosen arbitrarily. In
other words, the length of the current sheet $r_1$ is not
independent, which actually depends on the other two parameters,
$J$ and $h$. With the values of dimensionless current $J$, flux
rope major radius $h$, and the length of the current sheet $r_1$
fixed, all coefficients $a_{2k}$ and $d_{2i+1}$ are solved via
Equation (\ref{ES3}) and (\ref{ES4}). Once these coefficients are
determined, the global configuration of the magnetosphere is established.

\section{Catastrophic Loss of Equilibrium of Flux Ropes Induced by\\
Variations on Magnetar Surface}

In our catastrophic eruption model, rapid dynamical response of
the flux rope is initiated by the long timescale quasi-static
alterations at the magnetar surface \citep{HY14}.
The flux rope quasi-statically evolves in stable equilibrium states
until the critical loss of equilibrium point is
reached \citep{Yu12,YH13}. In this paper, we consider how a new
physical element, the current sheet connecting the magnetar surface and the flux rope ambience,
which would dramatically affect the flux rope's behavior.

\subsection{Physical Constraints for Equilibrium State}
The flux rope is considered to be in quasi-static equilibrium states.
Both the local internal equilibrium and the global external equilibrium 
constraints must be satisfied in order to keep the flux rope in equilibrium state \citep{Yu12}.
The internal equilibrium constraint indicates that the
force-free condition, $\mathbf{J}\times\mathbf{B}=0$, is valid
inside the flux rope. Following Paper I, the current density and
the magnetic field inside the flux rope is specified in terms of Lundquist's solution \citep{Lund50}.
In this case, the conservation in axial magnetic flux 
of the flux rope \citep{Yu12} implies that
$r_0$, the minor radius, is inversely proportional to $I$, the current carried by the flux rope.
Explicitly, $r_0=r_{00}I_0/I=r_{00}/J$, where
the dimensionless current $J$ is scaled by $I_0$ as $J\equiv
I/I_0$ and $r_{00}$ is the value of $r_{0}$ as $J=1$. Note that
$I_0$ is related to the magnetic flux $\Psi_0$ via $I_0 =\Psi_0
c/r_s$, where $c$ and $r_s$ are speed of light and magnetar
radius, respectively.

There are two requirements for the external equilibrium
constraint. The first requirement is the condition of force-balance. 
It is satisfied when the magnetic field induced by the current carried by 
the flux rope, $B_{\rm s}$ \citep{Shaf66}, which provides
an outward force, is balanced by the external magnetic field
$B_{\rm e}$, which provides an inward force. The second one is the
ideal frozen-flux condition. It demands the value of stream function on the flux rope edge 
$\Psi(h-r_0,0)$, keep constant in the evolution of magnetosphere.

The above mentioned two equilibrium constraints can be cast in the following forms: 
\citep{Yu12,HY14}
\begin{eqnarray}
    \left\{\begin{array}{ccccc}
        f(\sigma,J,h)&\equiv&B_{\rm s}-B_{\rm e}\ =\ 0 \\
        g(\sigma,J,h)&\equiv&\Psi(h-r_0,0)={\rm const} \\
    \end{array} \right. \ ,
\end{eqnarray}
where explicit forms of the two functions $f(\sigma,J,h)$ and
$g(\sigma,J,h)$ can be found in Paper I. For a given value of the
magnetic flux on magnetar surface $\sigma$, the current $J$, and
the height $h$ can be obtained by solving the above nonlinear
equation set for $J$ and $h$ via the Newton-Raphson method \citep{Press92}.

\subsection{Catastrophic State Transitions with Current Sheets}
In this section, we focus on the catastrophic behavior caused by
the flux injection from the interior of magnetar 
\citep{GH10,GMH07,Lyut06,TLK02}, while the crust gradual motion
will be studied separately \citep{Rude91,YHprep}. The surface
magnetic flux would gradually decrease if new magnetic fluxes with
opposite polarity are injected from the magnetar interior.
The flux rope would quasi-statically evolves in stable equilibrium states. 

\begin{figure}
\includegraphics[scale=0.7]{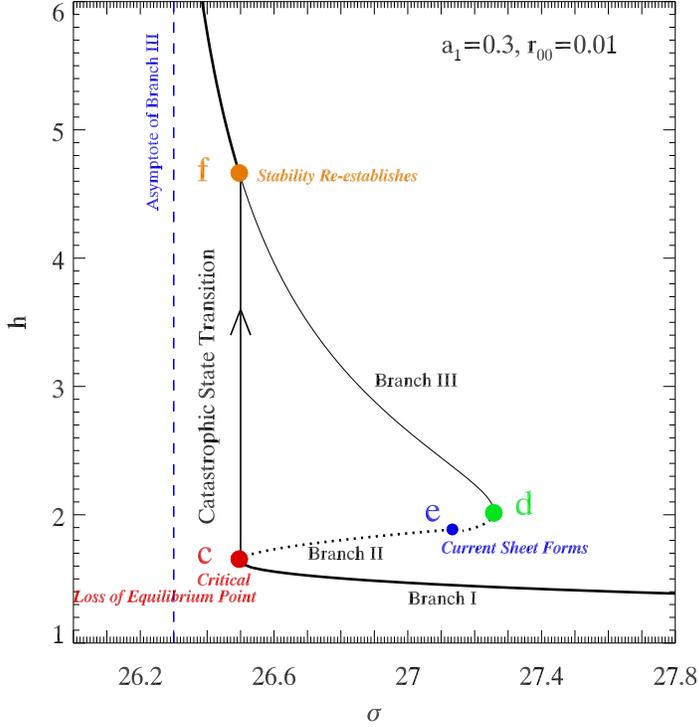}
\caption{\label{M3} The equilibrium curve for the catastrophic
state transition. The flux rope minor radius is $r_{00} = 0.01$
and the background field is a multipolar dominated field with
$a_1=0.3$. The equilibrium curve consists of three branches.
Branch I and III shown in solid curves are stable, while Branch II
shown in dotted curve is unstable. The current sheet starts to
form at the point $``e"$ (blue dot).
The flux rope undergoes a catastrophic state transition at the
critical loss of equilibrium point $``c"$. The flux rope transit to
the point $``f"$ (orange dot). Then it evolves quasi-statically
along Branch III above this point, approaching the vertical
asymptote (blue dashed line).
}
\end{figure}

We show an example of the equilibrium curve of the system 
in Fig.\ref{M3}, which describes the variation of the height of
flux rope $h$ with the surface magnetic flux of the central
neutron star $\sigma$. Estimation of \citet{TD01} suggests that
the typical radius of a flux rope is around $0.1$ $\mathrm{km}$. For
a magnetar with radius $10$ $\mathrm{km}$, this means that $r_{00}
\sim 0.01$. To be specific, we adopt $r_{00}=0.01$ throughout this
section. The background magnetic field, in which the flux rope is
embedded, is the multipolar dominated field specified by Equation
(\ref{pb}) with $a_1=0.3$. The equilibrium curve consists of three
branches. Branch I indicates the thick solid curve below the
critical loss of equilibrium point, denoted by a red dot $``c"$.
Branch II consists of the dotted curve between the red dot $``c"$
and the green dot $``d"$, and Branch III consists of the solid
curve above the point $``d"$. The current sheet starts to form at the point
represented by a blue dot $``e"$ in Branch II. Note that, when 
the height of flux rope is lower than the height where the current
sheet forms, $h_e \sim1.88 r_s$, the current sheet has not formed.
In this case, Region III covers the entire magnetosphere and the
global the magnetic field configurations can be obtained in the
same manner as in Paper I. We only need to specify the
coefficients $c_{2i+1}$ and $d_{2i+1}$ in Equation (\ref{psiIII})
according to the boundary condition 
to establish the global magnetic field configuration.
When the height of flux rope is larger than $h_e$, the current sheet starts to form. 
Then the global stream function should be
obtained from local solutions $\Psi_{\rm I}$, $\Psi_{\rm II}$, and
$\Psi_{\rm III}$ in different regions following the domain
decomposition method outlined in Section 3.

It is obvious that, when the current sheet is taken into account,
the equilibrium curves become more complicated than those obtained
in Paper I. In Fig.\ref{M3}, the two solid branches denote stable
branches. A flux rope lying on the these branches would behave as
a harmonic oscillator with slight displacements
\citep{Forb10,Yu12,HY14}. The dashed branch between the two solid
branches corresponds to an unstable branch. On this branch, a
slight deviation from the the equilibrium state would make the
flux rope depart from the equilibrium state further away \citep{Yu12}.
We consider a flux rope which stays on Branch I in the beginning,
i.e., the lower solid branch of the equilibrium curve.
In response to the gradually decreasing surface magnetic flux, 
the flux rope would rise up quasi-statically.
Once the flux rope evolves to the critical point (red point
$``c"$) with height $h_c \sim 1.64 r_s$, where Branch I and Branch II are connected, the flux rope
would lose its equilibrium and lead to violent eruptions. 
The flux rope jumps to a higher position, $h_f \sim 4.67 r_s$ catastrophically,
represented by an orange dot $``f"$ in Branch III. During this process of ``loss of equilibrium", 
The process of ``loss of equilibrium" would take place on a dynamical timescale, which is much shorter
than the quasi-static timescale of surface magnetic flux variations,
so that $\sigma$ could be regarded as unchanged.

It is interesting to note that the upper stable Branch III
approaches a vertical dashed asymptote in the $\sigma-h$
plane. Along this branch, the surface magnetic flux asymptotically
approaches  $\sigma_{\rm asym}$ in this plane. It is clear from
Fig.\ref{M3} that $\sigma_{\rm asym}< \sigma_c$. In such situations,
after the flux rope transits to the upper stable equilibrium state
at a higher position at $h=h_f$, the flux rope could re-enter the
stable equilibrium state and evolve along the thick segment on
Branch III and finally reach infinity on a quasi-static timescales. 
We call this behavior of the flux rope as the catastrophic state
transition, to distinguish it from the catastrophic escape
discussed in the following section.

\begin{figure}
\includegraphics[scale=1]{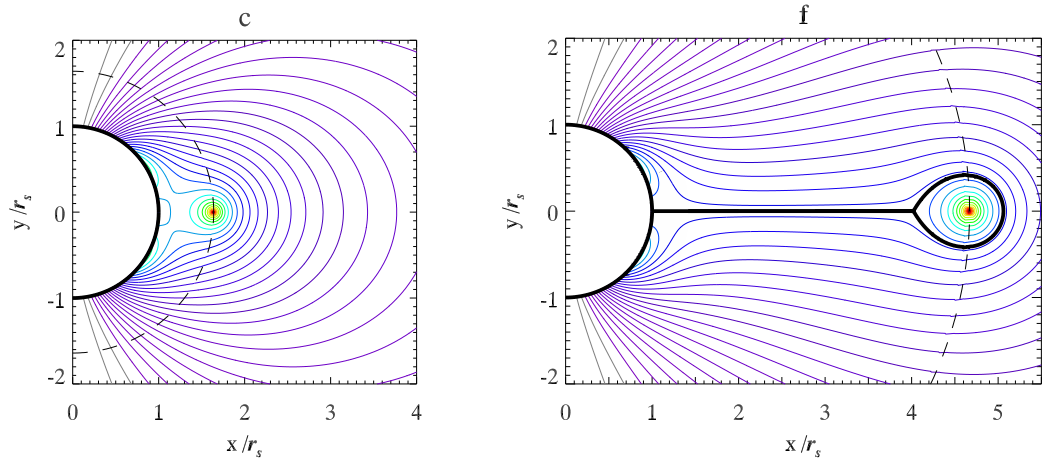}
\caption{\label{M3psi} {\it Left}: Configuration of the magnetic
field lines at the critical loss of equilibrium point $``c"$ shown
in Fig.\ref{M3}. The soild semi-circle represents the magnetar
surface. The dashed semi-circle represents the critical height of
the flux rope prior to catastrophic state transition. {\it Right}:
Configuration of the magnetic field lines at the point $``f"$ shown
in Fig.\ref{M3}. The thick horizontal solid line represents the
current sheet.
 }
\end{figure}

We show the magnetic field configuration corresponding to the
critical loss of equilibrium state $``c"$ in the left panel of
Fig.\ref{M3psi}. The thick solid semi-circle designates the
magnetar surface, and the thin dashed semi-circle with radius of $h_c$
presents the critical height of the flux rope prior to eruption.
In the right panel, 
the magnetic field configuration corresponding to the state $``f"$, where the stable equilibrium
state re-establishes, is presented.
The thin dashed semi-circle designates the height of $h_f$.
The current sheet is shown as a horizontal thick black line. 
The lower tip of the current sheet connects the magnetar surface. 
The upper tip of the current sheet connects flux rope by the Y-point at $r_1\sim4.0r_s$.
Huge magnetic energy will be released in short dynamical timescale
during the catastrophic state transition. The giant flare itself
may be well explained by such catastrophic state transitions.
However, after the catastrophic state transition, the flux rope
reaches another equilibrium branch at a higher position. Then it
evolves quasi-statically and such behavior is not consistent with
the radio afterglow associated with giant flares.

\subsection{Catastrophic Escape of Flux Ropes }
We have investigated the equilibrium curve of a flux rope in the
multipolar background field. The main differences from those in
Paper I is that an additional stable branch shows up in the
equilibrium curve when the current sheet formation is taken into
account. The state transition is triggered by the 
loss of equilibrium. With the re-establishment of stable
equilibrium state after the transition, the flux rope subsequent
evolution becomes quasi-static along the upper stable branch of
the equilibrium curve. Physically, this is because the magnetic
tension associated with the current sheet would counteract the
outward current-induced force during the 
evolution and prevent the flux rope from escaping.

The equilibrium curve of the flux rope depends on the background
field\footnote{Note that the flux rope's minor radius, $r_{00}$,
also has important implications for the flux rope's behavior when
the current sheet is considered. Actually, the bigger the minor
radius $r_{00}$ is, the easier for the flux rope to erupt. If the
value of $r_{00}$ is too small, the flux rope may not experience
the catastrophic eruption discussed here. To focus on the
catastrophic behavior 
we will defer our
discussion about the behavior of flux rope with different value of
minor radius, $r_{00}$, in Section 5.}.
Interestingly, we find that in certain background fields, a flux
rope with typical $r_{00}=0.01$ cannot reach new stable
equilibrium state once the flux rope evolves to the critical loss
of equilibrium point.
As a result, the flux rope would escape to infinity dynamically.
This behavior is different from the catastrophic transition we
mentioned above and we call it the catastrophic escape.

\begin{figure}
\includegraphics[scale=0.7]{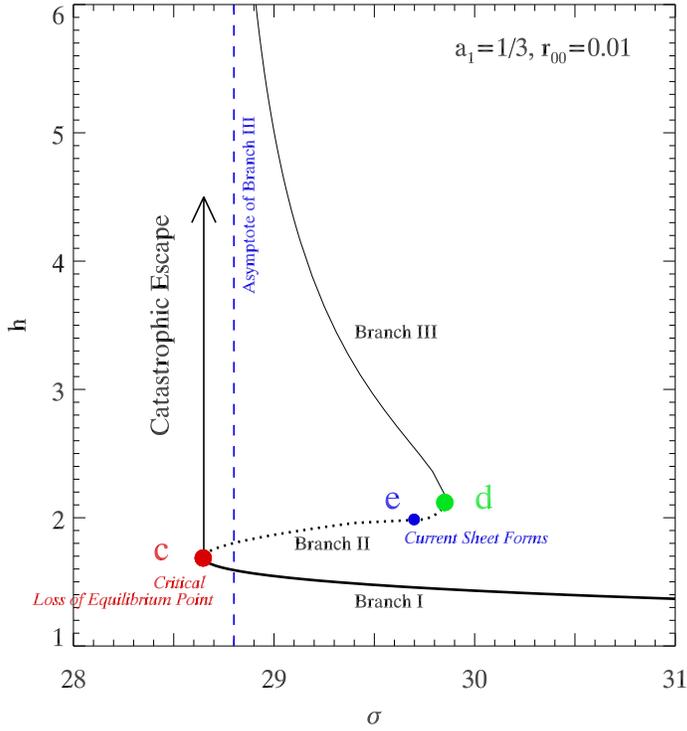}
\caption{\label{M33} The equilibrium curve for the catastrophic
escape. The background field is multipolar dominated with
$a_1=1/3$ and the flux rope minor radius is $r_{00}=0.01$. The
equilibrium curve consists of three branches. The asymptote (blue
dashed line) is located to the right of the critical loss of
equilibrium point $``c"$. Once the flux rope lose its equilibrium
at the point $``c"$, it would escape to infinity on a dynamical
timescale.
}
\end{figure}

As an example, we show the equilibrium curve of the flux rope in a
multipolar background field with $a_1=1/3$ and $r_{00}=0.01$ in
Fig.\ref{M33}. Similar to Fig.\ref{M3}, the equilibrium curve
includes three branches. The stable Branch I is shown as the thick
solid curve below the red dot $``c"$, the critical loss of
equilibrium point. The unstable Branch II is shown as the dotted
curve between the red dot $``c"$ and the green dot $``d"$. And the
stable Branch III is presented 
as the thin solid curve above the green
dot $``d"$. The current sheet starts to form at the blue dot $``e"$
on Branch II. The equilibrium curve below the point $``e"$
represents the evolution of the flux rope prior to the current
sheet formation, while the part above it represents the evolution
after the current sheet formation. This equilibrium curve also
presents an asymptotic behavior. The flux rope approaches 
$h\rightarrow \infty$ when surface magnetic flux $\sigma
\rightarrow\sigma_{\rm asym}$. The state transition would take
place only if the corresponding value of the surface magnetic flux
at the critical loss of equilibrium point $``c"$, $\sigma_c$, is
greater than $\sigma_{\rm asym}$ (see Fig.\ref{M3}.) Otherwise, as
shown in this example, the vertical asymptote indicates
$\sigma_{\rm asym}\approx28.8$ and at the red dot
$\sigma_c\approx28.65$.
It is interesting to note the fact that $\sigma_c < \sigma_{\rm asym}$,
which leads to the catastrophic escape behavior. The consequence is that, once the
system evolves to the critical point, the flux rope would
experience the eruption and jump upward. However, Branch III of
the equilibrium curve cannot be reached by the flux rope. The flux
rope cannot attain another stable equilibrium state but finally
escape to infinity.

During the catastrophic escape, the eruption of flux rope occurs
on a dynamical timescale. Huge magnetic energy will be released
during the flux rope eruption.
It is reasonable to expect that the flux rope's escaping to
infinity would correspond to the dynamical radio afterglow
associated with magnetar giant flares.
In the following section we will investigate in further detail the
energetics of the eruption of flux rope 
and how the multipolar background fields influence its catastrophic behavior.

\section{Energetics of Catastrophic Flux Rope Eruptions}
We have already known that once the flux rope evolves to the
critical point, it would lose its equilibrium and lead to an
eruption. When a current sheet appears during the catastrophic process, 
the flux rope would either transit to another stable equilibrium point at a
higher position or escape to infinity. In either type of
eruptions, magnetic energy would be released on a short dynamical
timescale. In this section, we will investigate the detailed
energetics of both the catastrophic transition and the
catastrophic escape of the flux rope.

\subsection{Energy Release with Catastrophic Transitions and Escapes}
Note that magnetic energy releases during the catastrophic process. 
The magnetic energy of magnetars is roughly
$E_{\rm mag} \sim 10^{48} (B/10^{15} {\rm G})(R/10 {\rm km})^3$
erg. The most powerful giant flare energy release is about $10^{46}$
erg. Approximately, only $1\%$ of the magnetic energy release is sufficient to drive giant flares.
It would be interesting to explore the energetics of the catastrophic eruption model. In our
model, the total magnetic energy of the magnetosphere can be
estimated by the following equation,
\begin{equation}
W_{t}(h) = - \int_{\infty}^h F(h^{\prime})\ d h^{\prime} + W_{\rm pot} \ ,
\end{equation}
where $F$ is the total magnetic force on the flux rope,
which can be calculated explicitly as $F=2 \pi I h (B_s-B_e)/c$.
It is clear that the first term in the above equation represents
the work required to move the flux rope from infinity to the place where the flux rope resides. 
The work integral is performed along the path with constant surface flux $\sigma$. 
Note that the external magnetic field, $B_e$, depends on the global field
configurations \citep{Yu12}. The second term, $W_{\rm pot}$, is
the magnetic potential energy of the background multipolar field.
The magnetic potential energy of the background multipolar field,
$W_\mathrm{pot}$, can be written as
\begin{equation}
\label{pot} W_\mathrm{pot} = \int \frac{{\bf B}_{\rm
pot}^2}{8\pi}\ dV = \int _{\partial V} \frac{B_{\rm pot}^2}{8\pi}
({\bf r} \cdot d{\bf S}) - \frac{1}{4\pi} \int_{\partial V} ({\bf
B}_{\rm pot}\cdot {\bf r}) ({\bf B}_{\rm pot}\cdot d{\bf S})\ .
\end{equation}
where the volume integral is performed over the entire magnetosphere. 
$\bf r$ is the position vector, and $d{\bf S}$ is the surface area element which directs outwards.
The potential magnetic field ${\bf B}_{\rm pot}$ is calculated 
by taking the spatial derivative of the potential stream function $\Psi_{\rm pot}$,
According to the boundary condition we adopted in Section 2,
it is explicitly written as $\Psi_{\rm pot}(r,\mu) =
\Psi_0\sigma_c\ \left[ (r_s/r) + a_1\ (5\mu^2-1) (r_s/r)^3
\right]\ (1-\mu^2)$.

Let's first focus on the energy release during the catastrophic
state transitions of the flux rope. As shown in Fig.\ref{M3}, the
magnetic energy release is determined by the difference in energy 
between two transition states, state $``c"$ and state $``f"$ in
Fig.\ref{M3}. The energy release fraction during state transitions
is $\Delta W_t/W_t(h_c)\equiv[W_t(h_c)- W_t(h_f)]/W_t(h_c)$.
Detailed calculation show that the magnetic energy release
fraction is $\Delta W_t/W_t(h_c) \approx 0.6\%$. It is obvious
that this energy release is approximately feasible to drive a
magnetar giant flare, which requires $\sim1\%$ of the magnetic
energy possessed in magnetar to be released.

The energy release during the catastrophic escape of flux rope is
similarly determined by the difference in energy 
between state $``c"$ and the state at infinity. The corresponding energy release
fraction is defined as $\Delta W_t/W_t(h_c)\equiv[W_t(h_c)-
W_t(h_f\to\infty)]/W_t(h_c)$, where $h_f$ is approaching infinity.
The magnetic energy release fraction for the case shown in
Fig.\ref{M33} is $\Delta W_t/W_t(h_c)\approx 0.9\%$. This energy
release fraction is in good agreement with the 1\% of energy
release required by magnetar observations. Note the energy of the
radio afterglow is much weaker than the giant flare, about $10^2-10^3$
times weaker than giant flares \citep{Gaen05}. Thus the
energy release during the catastrophic escape is enough for both
the giant flare and the magnetar radio afterglow. More
importantly, the dynamical escape of the flux rope corresponds
perfectly to the dynamical radio afterglow required by observations.

\subsection{Energy Release in Multipolar Background Fields}
\begin{figure}
\includegraphics[scale=1]{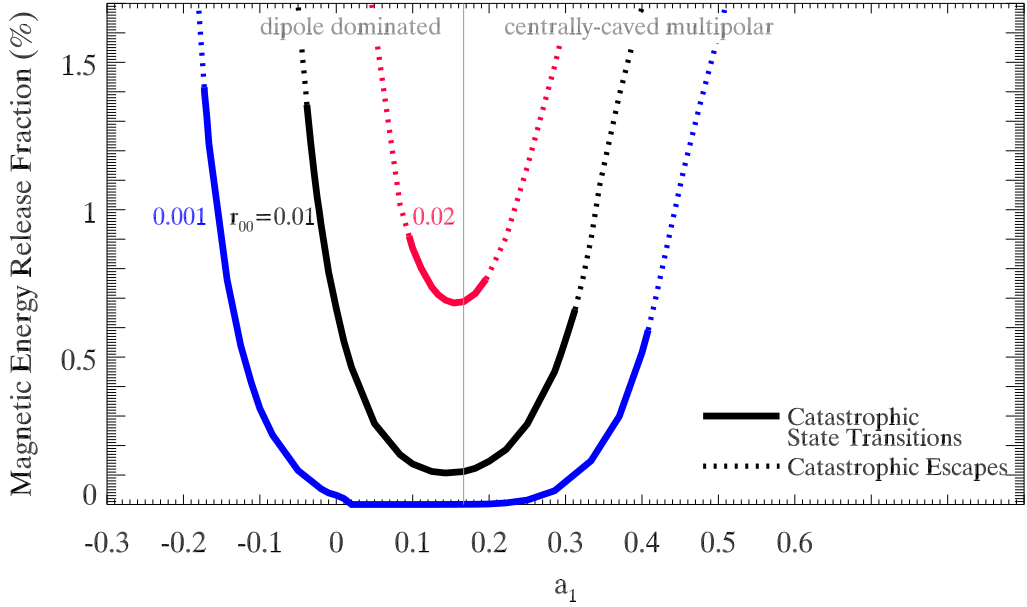}
\caption{\label{Wa1} Magnetic energy release fraction as a
function of background multipolar strengths $a_1$. Results for
flux ropes with minor radii $r_{00}=0.001$, $0.01$, and $0.02$ are
shown in blue, black, blue, and red lines, respectively. The solid
lines correspond to catastrophic state transitions and the dotted
lines correspond to catastrophic escapes. For flux ropes with
larger (smaller) minor radii, the value of $a_1$ for the
catastrophic state transition covers a narrower (wider) range in
$a_1$, which means that flux rope with larger minor radius is more
prone to experience the catastrophic escape. The catastrophic
escapes release more energy than the catastrophic state
transitions.
}
\end{figure}

Examples of the energy release for the catastrophic transitions
and catastrophic escapes have been mentioned above.
In this subsection, we will further study how the strength of
multipolar field affect the flux rope's catastrophic behaviors as
well as their energy release.
In Fig \ref{Wa1}, we show the variation of the energy release
fractions with the background multipolar field, the strength of
which is determined by the parameter $a_1$ in Equation (\ref{pb}).
Note that larger value of $|a_1|$ denotes stronger background
multipolar field. The blue, black, and red lines represent results
for flux rope minor radii $r_{00}=0.001$, $0.01$, and $0.02$,
respectively. Let us first focus on the black line, which
illustrates the dependence of energy release fractions on the
strength of the background multipolar field for a flux rope with
typical $r_{00}=0.01$. The solid part covers the range of
$-0.04\lesssim a_1\lesssim 0.32$, within which catastrophic state
transitions would take place.
This figure also indicates that approximately $0.1\%\sim1.35\%$ of
the magnetic energy in magnetosphere can be released for
catastrophic state transitions. The two dotted lines cover the
range $a_1\lesssim-0.04$ and $a_1\gtrsim0.32$, respectively,
within which the flux rope would experience catastrophic escapes.
It is obvious that the energy release associated with catastrophic
escapes is higher than the catastrophic state transition.

The flux ropes with different minor radii $r_{00}$ show similar
behaviors. According to Fig.\ref{Wa1}, the catastrophic state
transition for a flux rope with $r_{00}=0.001$ covers the range of
$-0.17 \lesssim a_1 \lesssim 0.41$ (the blue solid line). In the
range of $a_1\lesssim-0.17$ and $a_1\gtrsim0.41$ (the blue dotted
lines), the flux rope would experience catastrophic escapes. In
the regime of catastrophic state transitions,
the minimum of which is less than $10^{-3}\%$, while the maximum
of which is about $1.4\%$.
Catastrophic state transition of a flux rope with $r_{00}=0.02$
(the red solid line) could take place only for $0.095 \lesssim
a_1\lesssim0.2$. Outside this range, catastrophic escapes would
occur. The energy release fractions for $r_{00}=0.1$ is higher
than the case of $r_{00} = 0.001$ and $r_{00}=0.01$, the minimum
of which is about $0.7$ and the maximum of which is about $0.9\%$.
Interestingly, for flux ropes with smaller (larger) minor radii,
the value of $a_1$ for the catastrophic state transition to occur
covers a wider (narrower) range in $a_1$. In other words, a flux
rope with smaller minor radii is more difficult to escape from
central neutron star catastrophically.
Our calculation show that, if $r_{00}$ is larger than $\sim0.02$,
the flux rope would not re-attain a stable equilibrium state after
the catastrophic loss of equilibrium, and experience the
catastrophic escape only.

\subsection{Effects of Flux Rope Minor Radius}
The flux rope would experience either the catastrophic state
transition or the catastrophic escape, depending on the multipolar
field parameter $a_1$. In addition, we found that the flux rope
minor radius $r_{00}$ also plays an essential role for its
eruption. Here we would provide a further study on how the flux
rope minor radius influences its eruptions. For simplicity, we
consider a dipolar background field, i.e., $a_1=0$, in this
section.

\begin{figure}
\includegraphics[scale=0.6]{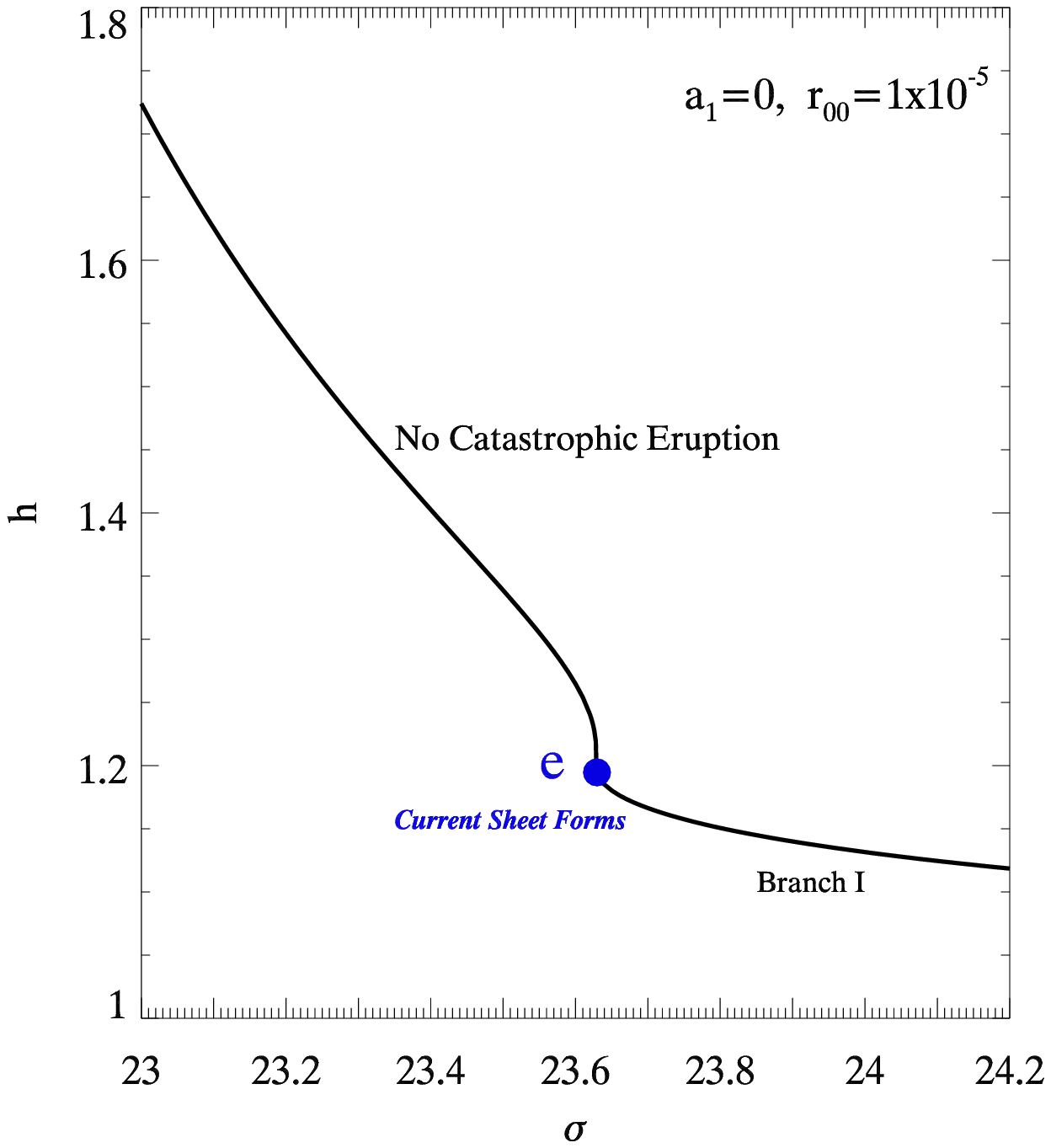}
\includegraphics[scale=0.6]{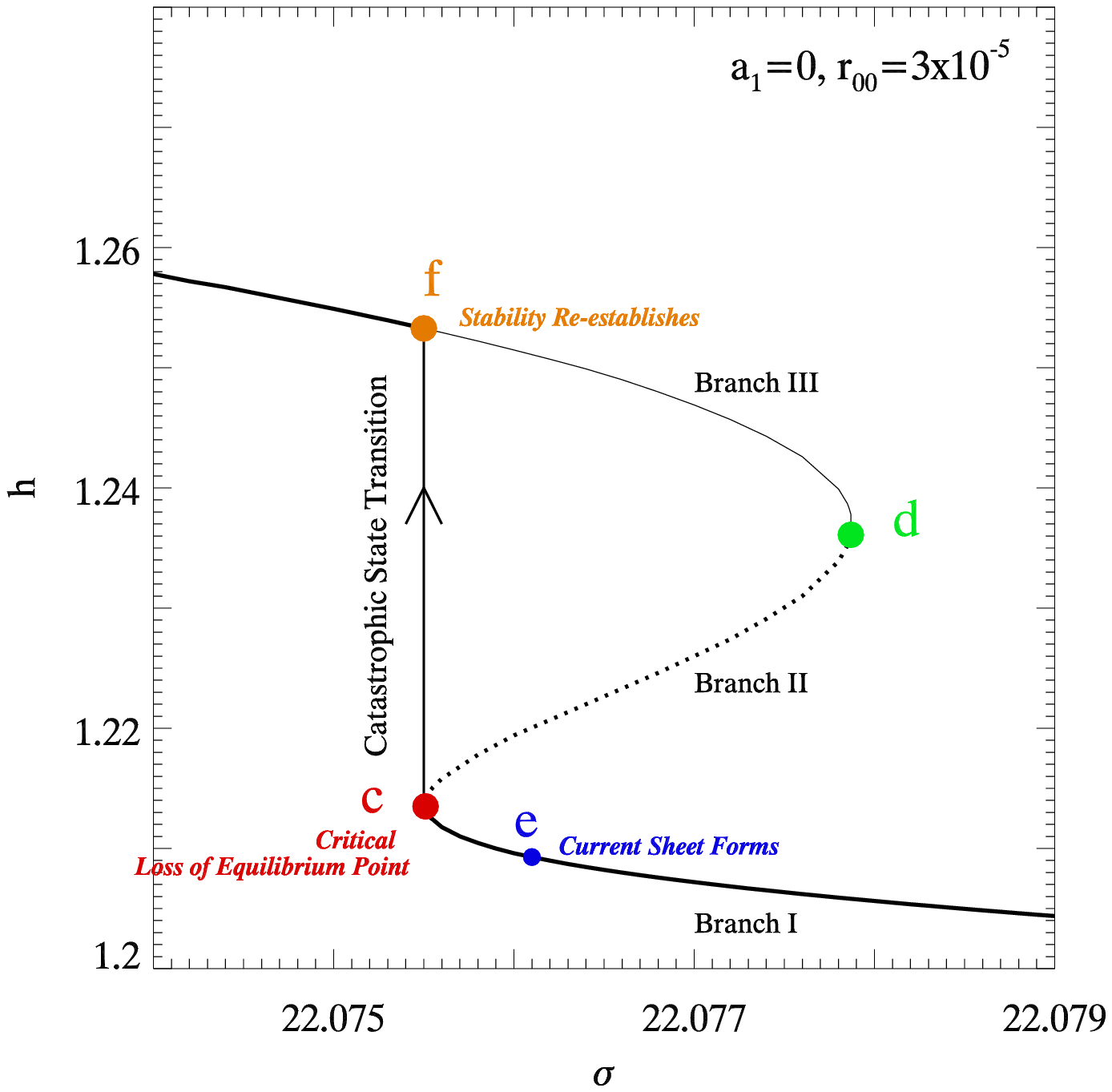}
\caption{ \label{M0r1-5} {\it Left}: Equilibrium curve of a flux
rope with $r_{00}=1\times 10^{-5}$ in dipolar background. The
equilibrium only has one stable branch and there is no
catastrophic eruption. {\it Right}: Equilibrium curve of a flux
rope with $r_{00}=3\times 10^{-5}$ in dipolar background. It
experiences a catastrophic state transition, similar to
Fig.\ref{M3}. The difference is that point  where current sheet
starts to form is located below the critical loss of equilibrium
point $``c"$.
}
\end{figure}

For different flux rope minor radius $r_{00}$, the equilibrium
curves may present new characteristics. Besides the catastrophic
state transition and the catastrophic escape, whose main features
are already shown in Fig.\ref{M3} and Fig.\ref{M33}, we find
alternative types of flux rope behavior. In the left panel of
Fig.\ref{M0r1-5}, we show the equilibrium curve for a flux rope
with $r_{00} = 1 \times 10^{-5}$. This equilibrium curve implies a
new flux rope behavior, which has only one stable branch and no
critical loss of equilibrium point. The current sheet starts to
form at the point $``e"$ and the magnetosphere evolves
quasi-statically all the way along the equilibrium curve without
eruption. In the right panel, the equilibrium curve is shown for
$r_{00} = 3 \times 10^{-5}$. Most features of this equilibrium
curve are similar to those shown in Fig.\ref{M3}, including the
three branches and the catastrophic state transition between the
points $``c"$ and $``f"$. The main difference is that the point
$``e"$, where the current sheet starts to form, is below the
critical loss of equilibrium point $``c"$. In other words, the
current sheet forms prior to the catastrophic eruption, this is
different from the flux rope behavior shown in Fig.\ref{M3}.

\begin{figure}
\includegraphics[scale=1]{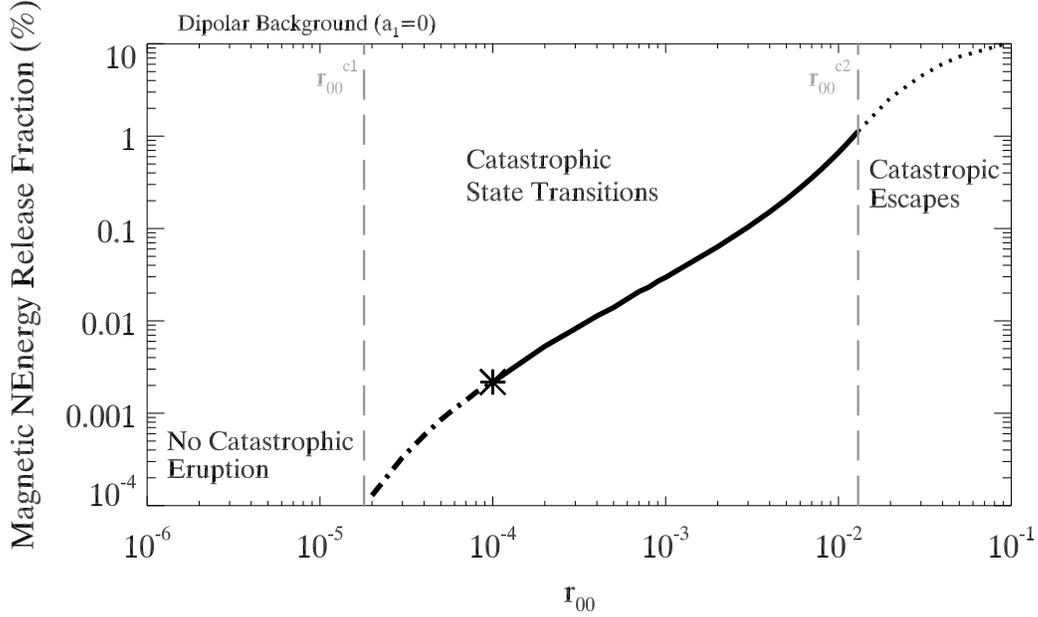}
\caption{ \label{Wr00} Dependence of magnetic energy release
fraction on the minor radius $r_{00}$ for a flux rope embedded in
a dipolar background. The flux rope undergoes catastrophic state
transitions provided the minor radius lies in the range
$[r_{00}^{c1}\approx1.8\times10^{-5},r_{00}^{c2}\approx0.013]$.
The two critical values of $r_{00}$ are shown by two vertical
dashed lines, respectively.
When the flux rope minor radius is small enough ($r_{00}<
r_{00}^{c1}$), the flux rope may not experience catastrophic
eruption. For flux rope with large enough radius ($r_{00}>
r_{00}^{c2}$), the catastrophic escape may occur (the dotted
line). Catastrophic state transition regime are shown by two
parts, the dot-dashed line (the current sheet forms prior to
catastrophic state transition) and the solid line (the current
sheet forms after catastrophic state transitions), separated by an
asterisk.
}
\end{figure}

The dependence of the magnetic energy release fraction on the flux
rope minor radius $r_{00}$ is given in Fig.\ref{Wr00}. We find
that the flux rope would undergo catastrophic state transition
provided its minor radius lies in the range of $[r_{00}^{c1},
r_{00}^{c2}]$, where the two critical minor radii are
$r_{00}^{c1}\approx 1.8\times10^{-5}$ and $r_{00}^{c2}\approx0.013$,
marked by two vertical dashed lines.
In the range with even larger minor radii, $r_{00} > r_{00}^{c2}$
(the dotted line), the catastrophic escapes would occur. 
Note that if the flux rope minor radius is small
enough, i.e., $r_{00} < r_{00}^{c1}$, it would evolve
quasi-statically without eruption. Note that there exist a
characteristic minor radius of $r_{00}^* \approx 10^{-4}$, denoted
by an asterisk in this figure. If the flux rope minor radius is
larger than this characteristic value, the current sheet formation
point $``e"$ is located above the critical loss of equilibrium
point $``c"$. In this case, the current sheet starts to form after
the system experience catastrophic state transitions. However, if
the flux rope minor radius is smaller than this value, the point
$``e"$ is located below the critical loss of equilibrium point
$``c"$ and the current sheet starts to from prior to the
catastrophic eruption.

\section{Conclusions and Disccusions}

We propose a catastrophic flux rope eruption model for the
magnetar giant flares as well as their associated radio
afterglows. The flux rope gradually evolves in stable equilibrium
states, in response to the quasi-static variations at the magnetar
surface. Upon the critical loss-of-equilibrium point is reached,
the flux rope is destabilized so that it erupts catastrophically.
In this paper, we extend our previous investigations about the
flux rope's physical behavior by incorporating a new physical
ingredient, a current sheet, into the force-free magnetosphere.
The topology of the magnetosphere is considerably changed by the
current sheet. To get the global configuration of the
magnetospheric field, a problem with mixed boundary value should be solved.
We figure out a domain decomposition method to deal with
the current sheet and we are able to get the equilibrium curve of the flux rope when the current sheet is
present in the magnetosphere.

We adopt multipolar boundary conditions to illustrate the observed complicated
geometry of magnetic field near the magnetar surface. 
When the current sheet appears, 
we find that the multipolar boundary conditions has influential implications
for the behavior of the flux rope. Depending on the strength of
the multipolar background field, the flux rope would experience
two types of catastrophic behaviors, the catastrophic state
transition and the catastrophic escape. In the catastrophic state
transition, the flux rope would jump to a new 
stable branch at a higher position and resume the quasi-static evolution.
The subsequent quasi-static behavior of
the catastrophic state transition is not consistent with radio
afterglow of giant flares. In the catastrophic escape, the flux
rope is not able to reach the stable equilibrium state and the
flux rope would move away from the magnetar dynamically. The
physical characteristic of the catastrophic escape is in good
agreement with the radio afterglow of giant flares.

In addition to the effects of multipolar background fields, we
also study how the flux rope minor radii influence its
catastrophic behavior. We find that it is much easier for flux
ropes with larger minor radii to erupt. If the minor radii are
sufficiently small, the flux rope would evolve quasi-statically
and could not undergo catastrophic eruption. This interesting
property may provide useful hint for the size of the flux rope on
magnetars.

Observationally, the total magnetic energy in the magnetosphere is
approximately $\sim 10^{48}({\rm B}/10^{15}G)^2(r_s/10{\rm km})^3$
ergs, and the maximum magnetic energy released during a giant
flare is about $\sim10^{46}$ergs, i.e., only about $1\%$ of the
total magnetic energy in the magnetosphere could account for a
giant flare. We carefully investigate the detailed energetics of
both the catastrophic state transition and the catastrophic
escape. We find that, with appropriate boundary condition, the
magnetic energy release of catastrophic state transitions can
reach the level of $1\%$, which is sufficient to support the
observed giant flares. The energy release for the catastrophic
escape is even larger than the catastrophic state transition. The
most attractive feature about the catastrophic escape is that that
dynamical escape of the flux rope may well correspond to the
dynamical radio afterglow observed for some magnetars. The
catastrophic escape of flux rope may be a better mechanism to
explain the giant flare and the associated radio afterglow.

It should be pointed out that
a good place for magnetic reconnection is provided in the location of the current sheet,
which showed up during the catastrophic eruption. The energy release would be further enhanced
if sufficiently fast magnetic reconnection proceeds in the current sheet.
The effects of magnetic reconnection can be further studied
by a time-dependent flux rope eruption model, in which magnetic
reconnection can be treated self-consistently.
Now we are trying to investigate the effects of magnetic
reconnection by studying the time-dependent behavior the flux rope
eruption. Further results about how magnetic reconnection
influences its eruptive behavior will be reported elsewhere \citep{YHprep}.

\acknowledgments This work has been supported by National Natural
Science Foundation of China (Grants 11203055, 11173057,
11373064, 11121062 and 11173046), Open Research Program in Key Lab
for the Structure and Evolution of Celestial Objects (Grant
OP201301), Yunnan Natural Science Foundation (Grant 2012FB187).
This work is partly supported by the Strategic Priority Research
Program ``The Emergence of Cosmological Structures" of the Chinese
Academy of Sciences (Grant No. XDB09000000) and the CAS/SAFEA
International Partnership Program for Creative Research Teams.
Part of the computation is performed at HPC Center, Yunnan
Observatories, CAS, China.


\appendix
\section{Numerical Matching for Different Regions}
The global magnetosphere is divided into three different
computational domains. The general solution in Region III
containing the flux rope has a form different from that in Region
I and II. Global solutions should make the magnetic fields in
Region III smoothly connected with those in Region I and II. In
this Appendix, we will show detailed procedures to construct such
global solutions. By taking the spatial derivatives of stream
functions in Region I and Region III,
we know that the matching conditions, Equations (\ref{matchBr})
and (\ref{matchBth}) can be written explicitly as follows,
\[
- \Psi_0\sigma\Theta(0) - \sum_{k=1}^\infty\left[
a_{2k}r_1^{2k+1}+(T_{2k}-a_{2k})r_1^{-2k} \right] \ P_{2k}(\mu)
\]
\begin{equation}\label{ES1}
= \ - \sum_{i=0}^\infty \left( c_{2i+1}\frac{r_1^{2i+2}}{h^{2i+2}}
+ \frac{d_{2i+1}}{ r_1^{2i+1}} \right)\ P_{2i+1}(\mu)\ ,
\end{equation}
and
\begin{eqnarray}
\label{ES2}
    &\ & \sum_{k=1}^\infty \left[ (2k+1)a_{2k}r_1^{2k}-(2k)(T_{2k}-a_{2k})r_1^{-2k-1} \right] \left[ \frac{P_{2k-1}(\mu) - P_{2k+1}(\mu)}{(4k+1)} \right]  \nonumber \\
    &=& \sum_{i=0}^\infty \left[ (2i+2)\frac{c_{2i+1}}{h} \left(\frac{r_1}{h}\right)^{2i+1}-(2i+1) \frac{d_{2i+1}}{r_1^{2i+2}} \right] \ \left[ \frac{P_{2i}(\mu) - P_{2i+2}(\mu)}{(4i+3)} \right]\ ,
\end{eqnarray}
where $\mu\in[0,1]$ and the relation $b_{2k}=T_{2k}-a_{2k}$
has been exploited.

Note that in the above two equations, 
Legendre polynomials with even orders and odd orders
appear separately on the two sides of the
equations. To make the two sides comparable, we need to expand
odd-order Legendre polynomials in terms of Legendre polynomials with even orders.
The odd-order Legendre, $P_{2i+1}(\mu)$, on the right
hand side of Equation (\ref{ES1}) can be expanded as
\begin{equation}
P_{2i+1}(\mu) = \sum_{k=0}^\infty (4k+1)C_{(2i+1)(2k)}\
P_{2k}(\mu) \ ,
\end{equation}
where
\begin{equation}
C_{(2i+1)(2k)} = \int_0^1P_{2i+1}(\mu)P_{2k}(\mu)\ d\mu  \ .
\end{equation}
For numerical convenience, we pre-calculate all the coefficients
$C_{(2i+1)(2k)}$ and store them for later use. When we substitute
the above expression for $P_{2i+1}(\mu)$ into the right hand side
of Equation (\ref{ES1}), we can get
\begin{eqnarray}
\label{ES1a}
    &\ & - \sum_{i=0}^\infty \left( c_{2i+1}\frac{r_1^{2i+2}}{h^{2i+2}} + \frac{d_{2i+1}}{ r_1^{2i+1}} \right)\ P_{2i+1}(\mu)  \nonumber \\
    &=& - \sum_{i=0}^\infty \left( c_{2i+1}\frac{r_1^{2i+2}}{h^{2i+2}} + \frac{d_{2i+1}}{ r_1^{2i+1}} \right)\ \left\{\ \sum_{k=0}^\infty (4k+1)C_{(2i+1)(2k)}\ P_{2k}(\mu)\ \right\}  \nonumber \\
    &=& - \sum_{k=0}^\infty \left\{\ \left[(4k+1)\sum_{i=0}^\infty C_{(2i+1)(2k)}c_{2i+1}\frac{r_1^{2i+2}}{h^{2i+2}} + (4k+1)\sum_{i=0}^\infty C_{(2i+1)(2k)} \frac{d_{2i+1}}{r_1^{2i+1}}\right]\ P_{2k}(\mu)\ \right\}\ .  \nonumber \\
\end{eqnarray}
Obviously, the left hand side of Equation (\ref{ES1}) can be
written as (since $P_0(\mu)\equiv1$),
\begin{eqnarray}
\label{ES1b}
- \Psi_0\sigma\Theta(0)P_0(\mu) - \sum_{k=1}^\infty
\left\{\ \left[\left( r_1^{2k+1} - \frac{1}{r_1^{2k}}
\right)a_{2k} + \frac{T_{2k}}{r_1^{2k}}\right]\ P_{2k}(\mu) \
\right\}\ .
\end{eqnarray}
Note that in Equation (\ref{ES1a}) and (\ref{ES1b}), coefficients
$c_{2i+1}$'s and $T_{2k}$'s are known quantities, which are
already determined according to $J$ and $\sigma$, respectively
\citep{Yu12}. While $a_{2k}$'s and $d_{2i+1}$'s are unknown
variables, which need to be specified according to the matching
conditions.
By comparing coefficients in front of each $P_{2k}(\mu)$ for
$k=0,1,\cdots,N-1$\footnote{Here $N$ is sufficiently large
integer. Typically, we choose $N=400$ for the Legendre expansions
in our calculations.} in Equation (\ref{ES1a}) and Equation
(\ref{ES1b}), we find that the unknown coefficients $a_{2k}$'s and
$d_{2i+1}$'s must satisfy the following linear equations
\begin{eqnarray}
\label{ES3}
    \sum_{i=0}^{N-1} C_{(2i+1)(0)}\ \frac{d_{2i+1}}{r_1^{2i+1}} &=& G_0 \ , \ \mathrm{for} \ k = 0 , \nonumber \\
    S_{2k} \ a_{2k} + \sum_{i=0}^{N-1} (4k+1)C_{(2i+1)(2k)}\ \frac{d_{2i+1}}{r_1^{2i+1}} &=& G_{2k}\ , \ \mathrm{for} \ k=1,\cdots,N-1 \ ,
\end{eqnarray}
where
\begin{eqnarray}
    G_0 &=& \Psi_0\sigma\Theta(0)\ -\ \sum_{l=0}^N C_{(2l+1)(0)}c_{2l+1}\frac{r_1^{2l+2}}{h^{2l+2}} \ ,  \nonumber \\
    S_{2k} &=& - \left( r_1^{2k+1} - \frac{1}{r_1^{2k}} \right) \ , \ \mathrm{for} \ k=1,\cdots,N-1 \ ,  \nonumber \\
    G_{2k} &=& \frac{T_{2k}}{r_1^{2k}} - (4k+1)\sum_{l=0}^N C_{(2l+1)(2k)}c_{2l+1}\frac{r_1^{2l+2}}{h^{2l+2}}\ , \ \mathrm{for} \ k=1,\cdots,N-1 \ . \nonumber
\end{eqnarray}
Note that there are totally  $N$ number of linear equations for
$a_{2k}$'s and $d_{2i+1}$'s in Equation (\ref{ES3}).

It is clear that another $N$ number of linear equations are
necessary to get all the unknown coefficients $a_{2k}$'s and
$d_{2i+1}$'s. These additional equations can be obtained from the
second matching condition, Equation (\ref{ES2}), with similar
manipulations described above.
The odd-order polynomials $P_{2k-1}(\mu)$ and $P_{2k+1}(\mu)$ can
be expanded in terms of Legendre polynomials with even orders as (we set
$s=k-1$),
\begin{eqnarray}
    P_{2k-1}(\mu) &=& P_{2s+1}(\mu) = \sum_{i=0}^\infty (4i+1)C_{(2s+1)(2i)}\ P_{2i}(\mu) \ , \nonumber\\
    P_{2k+1}(\mu) &=& P_{2s+3}(\mu) = \sum_{i=0}^\infty (4i+1)C_{(2s+3)(2i)}\ P_{2i}(\mu) \ .
\end{eqnarray}
Similar to the manipulations with Equation (\ref{ES1}), we
substitute the above expressions into the left hand side of
Equation (\ref{ES2}). The left hand side of Equation (\ref{ES2})
can be expanded in terms of Legendre polynomials with even orders as
follows,
\begin{eqnarray}
\label{ES2a}
    &\ & \sum_{s=0}^\infty \left[ (2s+3)a_{2s+2}r_1^{2s+2} - (2s+2)(T_{2s+2}-a_{2s+2})r_1^{-2s-3} \right] \left[ \frac{P_{2s+1}(\mu) - P_{2s+3}(\mu)}{(4s+5)} \right]  \nonumber \\
    &=& \sum_{i=0}^\infty \left\{\ \left[(4i+1)\sum_{s=0}^\infty \left( (2s+3)r_1^{2s+2} + \frac{(2s+2)}{r_1^{2s+3}} \right) \widetilde{A}_{(2s+1)(2i)}\ a_{2s+2} \right.\right.  \nonumber \\
    &\ & \left.\left. - (4i+1)\sum_{s=0}^\infty (2s+2)\widetilde{A}_{(2s+1)(2i)} \frac{T_{2s+2}}{r_1^{2s+3}}  \right]\ P_{2i}(\mu)\ \right\}\ ,
\end{eqnarray}
where
\begin{eqnarray}
    \widetilde{A}_{(2s+1)(2i)} &=& \frac{C_{(2s+1)(2i)} - C_{(2s+3)(2i)} }{(4s+5)}\ .
\end{eqnarray}
Similar to the treatment of $C_{(2i+1)(2k)}$ in Equation (A4), we
also pre-calculate all the coefficients
$\widetilde{A}_{(2s+1)(2i)}$ and store them for later use. The
left hand side of Equation (\ref{ES2}) is re-arranged as summation
of even-order Legendre polynomials $P_{2i}(\mu)$.
By matching coefficients in front of each Legenfre polynomial
$P_{2i}(\mu)$ for $i=0,1,...,N-1$ in Equation (\ref{ES2a}) and in
the right hand side of Equation (\ref{ES2}),
\[
\sum_{k=1}^{N} \left[ (2k+1)r_1^{2k} + \frac{2k}{r_1^{2k+1}}
\right] \widetilde{A}_{(2k-1)(0)}\ a_{2k}\ +\ \frac{1}{3}\
\frac{d_{1}}{r_1^{2}} = Q_0 \ , \hspace{10mm} \ \mathrm{for} \
i=0,
\]
\[
\sum_{k=1}^{N} (4i+1)\left[ (2k+1)r_1^{2k} + \frac{2k}{r_1^{2k+1}}
\right] \widetilde{A}_{(2k-1)(2i)}\ a_{2k}\ +
\frac{(2i+1)}{(4i+3)}\ \frac{d_{2i+1}}{r_1^{2i+2}}\ -
\frac{(2i-1)}{(4i-1)}\ \frac{d_{2i-1}}{r_1^{2i}} = Q_{2i} \ ,
\]
\begin{equation} \label{ES4}
\hspace{125mm} \ \mathrm{for} \  i=1,\cdots,N-1,
\end{equation}
where
\begin{eqnarray}
Q_0 &=& \frac{2c_1}{3h}\frac{r_1}{h} + \sum_{m=1}^N 2m\widetilde{A}_{(2m-1)(0)} \frac{T_{2m}}{r_1^{2m+1}}\ ,  \nonumber \\
Q_{2i} &=&
\frac{(2i+2)}{(4i+3)}\frac{c_{2i+1}}{h}\left(\frac{r_1}{h}\right)^{2i+1}
-
\frac{2i}{(4i-1)}\frac{c_{2i-1}}{h}\left(\frac{r_1}{h}\right)^{2i-1}\
+ (4i+1)\sum_{m=1}^N 2m\widetilde{A}_{(2m-1)(2i)}
\frac{T_{2m}}{r_1^{2m+1}}\ ,  \nonumber
\end{eqnarray}
Equation (\ref{ES4}), together with Equation (\ref{ES3}), provides
$2N$ number of linear equations for $2N$ number of unknown
variables, i.e., $N$ number of unknowns $a_{2k}$ with
$k=1,2,...,N$ and $N$ number of unknowns $d_{2i+1}$ with
$i=0,1,2,...,N-1$. These unknowns can be solved numerically by a
standard linear solver. Once these unknown coefficients are fixed,
the global magnetospheric field with a current sheet can be
established.




\end{document}